\documentclass[floats,floatfix,showpacs,amssymb,prd,twocolumn,superscriptaddress,nofootinbib,nolongbibliography,reprint]{revtex4-2}

\usepackage{amssymb,amsmath,verbatim,mathtools,needspace,enumitem,etoolbox,graphicx,physics,microtype,afterpage,bigints,gensymb,tabularx}

\usepackage[dvipsnames, usenames]{xcolor}
\definecolor{linkcolor}{rgb}{0.0,0.3,0.5}
\definecolor{dodgerblue}{HTML}{1E90FF}
\usepackage[unicode, colorlinks=true, linkcolor=linkcolor, citecolor=linkcolor, filecolor=linkcolor,urlcolor=linkcolor, pdfusetitle]{hyperref}
\usepackage[all]{hypcap}
\usepackage[T1]{fontenc}
\usepackage[utf8]{inputenc}
\usepackage[normalem]{ulem}
\makeatletter
\newcommand*{\balancecolsandclearpage}{\close@column@grid \cleardoublepage \twocolumngrid}
\makeatother

\newcommand\orcid[1]{\href{https://orcid.org/#1}{$\!$\includegraphics[scale=0.006]{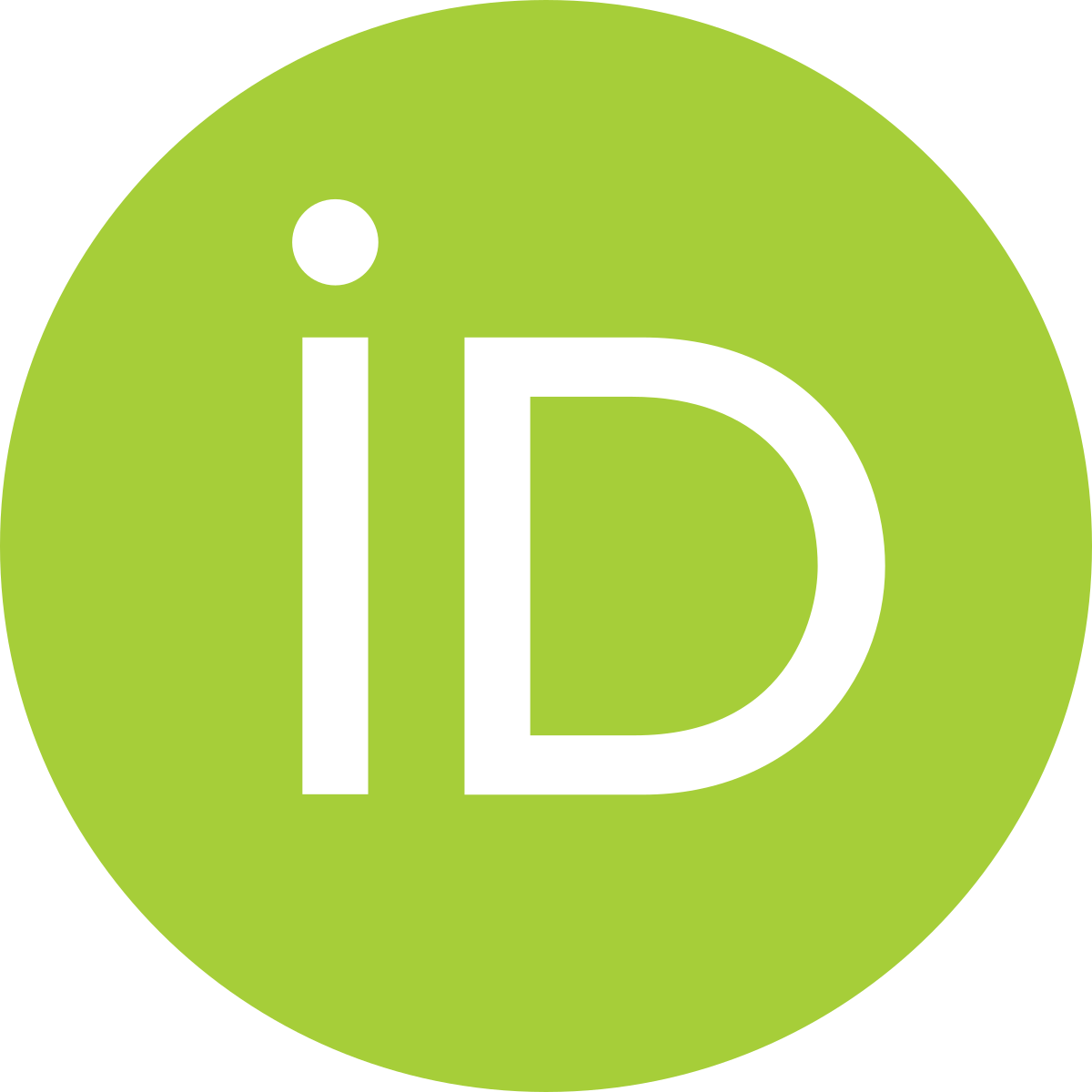} $\!\!$}}
\usepackage{lipsum}

\def\apj{\rm{ApJ}}
\def\apjl{\rm{ApJ}}
\def\apjs{\rm{ApJS}}

\def\mnras{\rm{MNRAS}}
\def\prd{\rm{PRD}}
\def\prl{\rm{PRL}}
\def\prx{\rm{PRX}}

\def\cqg{\rm{CQG}}

\newcommand{\bham}{\affiliation{School of Physics and Astronomy \& Institute for Gravitational Wave Astronomy, \\ University of Birmingham, Birmingham, B15 2TT, United Kingdom}}

\newcommand{\milan}{
\affiliation{Dipartimento di Fisica ``G. Occhialini'', Universit\'a degli Studi di Milano-Bicocca, Piazza della Scienza 3, 20126 Milano, Italy}
\affiliation{INFN, Sezione di Milano-Bicocca, Piazza della Scienza 3, 20126 Milano, Italy}
}

\begin{document}

\title{Gravitational-wave population inference at past time infinity}
\author{Matthew Mould \orcid{0000-0001-5460-2910}}
\email{mmould@star.sr.bham.ac.uk}
\bham

\author{Davide Gerosa \orcid{0000-0002-0933-3579}}
\email{davide.gerosa@unimib.it}
\milan \bham

\pacs{}

\date{\today}

\begin{abstract}
Population studies of stellar-mass black hole binaries have become major players in gravitational-wave astronomy. The underlying assumptions are that the targeted source parameters refer to the same quantities for all events in the catalog and are included when modeling selection effects. Both these points have so far been neglected when estimating the orientations of the black hole spins. In particular, the detector-frame gravitational-wave frequency used to define frequency-dependent quantities (e.g., 20 Hz) introduces an inconsistent reference between events at the population level. We solve both  issues by modeling binary black hole populations and selection effects at past time infinity, corresponding to the well-defined reference frequency of 0 Hz. We show that, while current gravitational-wave measurement uncertainties obfuscate the influence of reference frequency in population inference, ignoring spins when estimating selection effects leads to an overprediction of spin alignment in the underlying astrophysical distribution of merging black holes.

\end{abstract}

\maketitle

\section{Introduction}

The population of stellar-mass black hole (BH) binaries detected via gravitational waves (GWs) is growing quickly \cite{2019PhRvX...9c1040A,2021PhRvX..11b1053A}. A key underlying assumption when performing a statistical analysis at the population level is that one is combining measurements of the \emph{same} parameters for all events. This is trivially the case for constant quantities such as masses and spin magnitudes. The spin orientations, however, are subject to relativistic precession and vary as the binary inspirals toward merger~\cite{1994PhRvD..49.6274A}. If present, the orbital eccentricity also changes during the inspiral~\cite{1963PhRv..131..435P}.

State-of-the art GW population inference is performed at a fixed detector-frame frequency~\cite{2021ApJ...913L...7A}. For LIGO/Virgo, the reference frequency is typically set to $f_{\rm ref} = 20$ Hz. This choice is perfectly acceptable when analyzing a single event but questionable at the population level, mainly for two reasons:
\begin{itemize}
\item As currently defined, $f_{\rm ref}$ is a detector-frame quantity, but the parameters targeted in the inference (e.g., masses) are taken in the source frame.
\item The binary configuration at a given reference frequency depends on the other parameters. Systems with larger (smaller) masses and/or antialigned (coaligned) spins are closer (farther) from merger.
\end{itemize}
These issues were made evident for GW190521, where the high mass imposed a lower value $f_{\rm ref}=11$~Hz, different from all other events in the catalog~\cite{2021PhRvX..11b1053A}.

The spin directions at $f_{\rm ref}$ thus describe different quantities for each event: one is \emph{not} allowed to put them together in a population fit as if they were the same. To the best of our knowledge, this issue plagues all current GW population studies which make use of quantities that vary during the binary evolution.

A first step in the right direction was recently taken in Ref.~\cite{2021arXiv210709692V}, where spins are quoted at a fixed dimensionless time $t_{\rm ref}=-100 M$ from the peak of the GW strain. While this tackles the first issue highlighted above, it does not fully address the second point. One can still construct several dimensionless quantities that vary monotonically along the inspiral (e.g., time, orbital frequency, orbital separation) and select any of those when quoting the spin directions.

Current spin inference suffers from another pressing issue. When reconstructing the \emph{observable} population of sources from the \emph{observed} catalog, one must account for selection effects~\cite{2019MNRAS.486.1086M,2020arXiv200705579V}. Although it is well known that sources with spins coaligned (counteraligned) with the binary's orbital angular momentum are easier (harder) to observe~\cite{2001PhRvD..64l4013D,2018PhRvD..98h3007N}, this detection bias is often deemed as unimportant and neglected. While (the aligned components of) the spins are included in the pipeline injections used to estimate the LIGO/Virgo detectability, their dependence on the population parameters is then neglected when sampling the population likelihood~\cite{2021ApJ...913L...7A}.

We present the first complete solution to these conceptual issues: the spin directions needed in both the population and selection-effect models are those at past time infinity or, equivalently, $f_{\rm ref}=0$ Hz. This asymptotic configuration uniquely determines the entire history of the binary up to an orbital and a precessional phase~\cite{2015PhRvD..92f4016G,2020CQGra..37v5005R}. This reference puts the spin directions of all the events on equal footing, thus allowing for a consistent implementation of the population fit.
Two points need to be tackled:
\begin{itemize}
\item Data from GW events are, by definition, taken when the binary is detectable and thus need to be propagated backward to past time infinity.
\item Modeling selection effects instead requires propagating the tested population forward from past time infinity to detection.
\end{itemize}
We first travel back in time (20 $\to$ 0~Hz) when treating the event likelihoods and then ``back to the future'' (0 $\to$ 20~Hz) when handling selection effects. Our DeLorean consists of precession-averaged post-Newtonian (PN) evolutions.

\begin{figure*}
\includegraphics[scale=0.55]{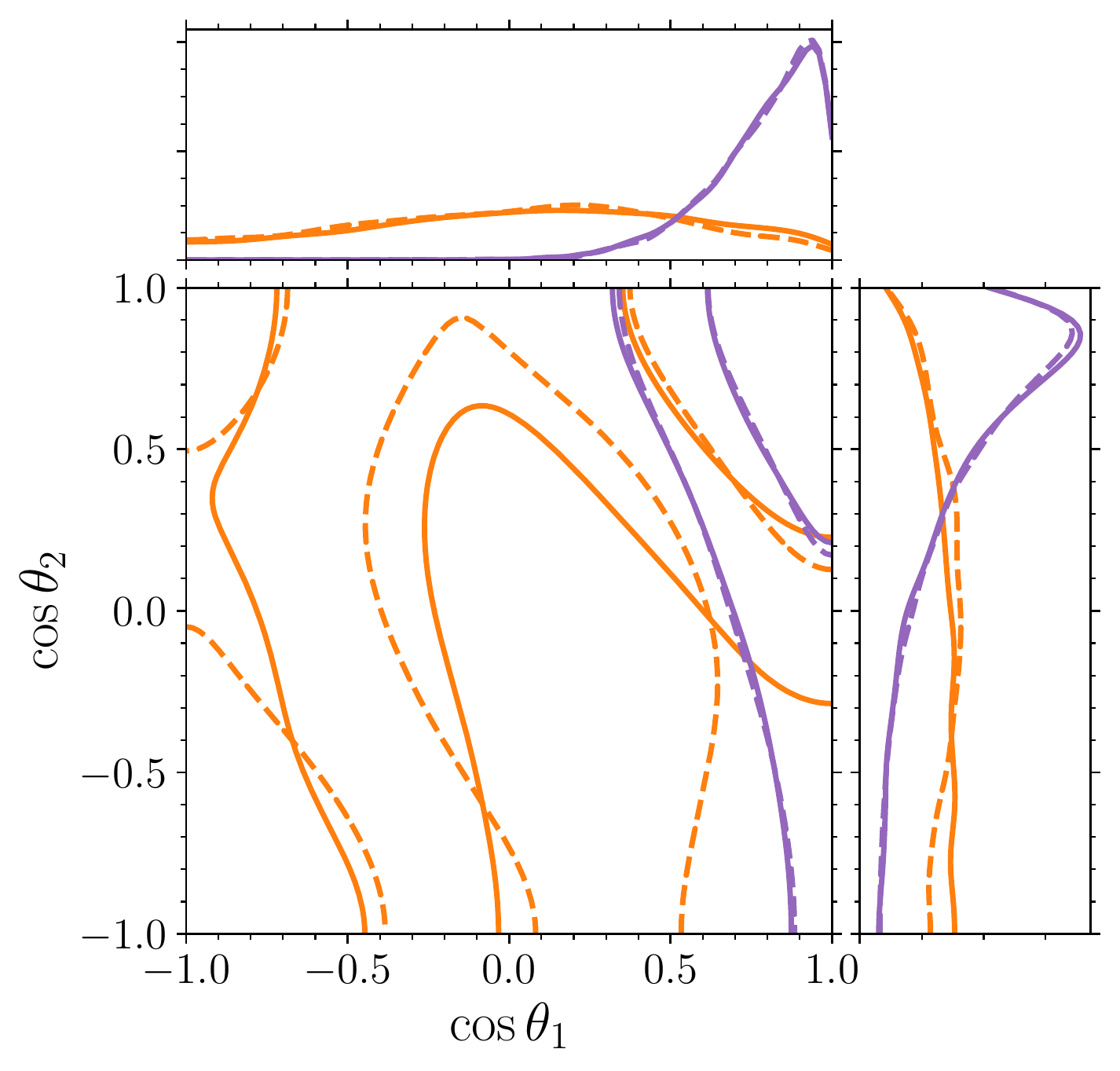}
$\quad$
\includegraphics[scale=0.55]{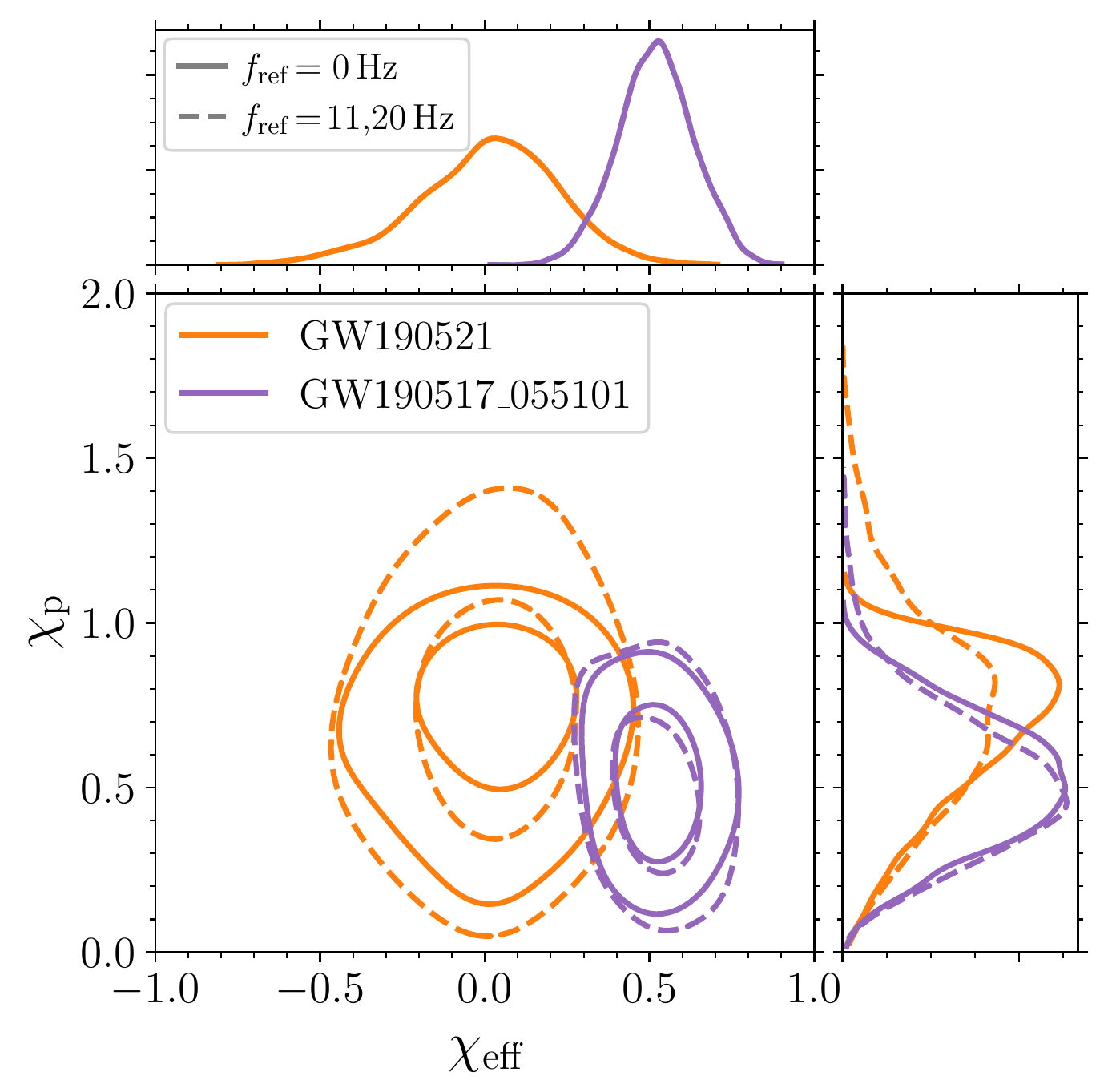}
\caption{Back-propagating GW data to past time infinity. We show the event with the largest mass (GW190521, orange) and that with the largest effective spin (GW190517\_0555101, green). Left and right panels show the joint posterior distribution of the spin tilts $(\theta_{1},\theta_{2})$ and the effective spins $(\chi_{\rm eff},\chi_{\rm p})$, respectively. Dashed curves show the distributions in the LIGO/Virgo band ($f_{\rm ref}=11$ Hz for GW190521 and $f_{\rm ref}=20$ Hz for GW190517\_0555101). Solid curves are computed at  $f_{\rm ref}=0$ Hz. The central panels show contours at the 50\% and 90\% levels.}
\label{backpropfig}
\end{figure*}

\section{Statistical inference}

The statistical problem we tackle is that of an inhomogeneous Poisson process including  measurement errors and selection effects \cite{2004AIPC..735..195L,2019MNRAS.486.1086M,2020arXiv200705579V}. We denote the parameters of individual events with $\theta$  (e.g. BH masses, spins, etc.) and those of the overarching population with $\lambda$ (e.g. power-law index of the mass spectrum, etc.). The targeted posterior is
\begin{equation}
\label{poppost}
p(\lambda | d) \propto \pi(\lambda) \sigma^{-N}(\lambda) \prod_{i=1}^{N}  \int p_{\rm pop}(\theta|\lambda) \mathcal{L}({d_i}|\theta)\, {\rm d} \theta
\end{equation}
where $i=1,...,N$ labels the events in the catalog, $d$ indicates the entire data stream, $d_i$ indicates a short stretch of data around event $i$, $\mathcal{L}({d_i}|\theta)$ is the likelihood of the single-event analysis, $p_{\rm pop}(\theta|\lambda)$ is the population model, and $\pi(\lambda)$ is a prior on the population parameters. Selection effects enter the population likelihood via
\begin{equation}
\label{eqseleff}
\sigma(\lambda) = \int p_\mathrm{pop}(\theta|\lambda) p_\mathrm{det}(\theta) \dd\theta
\end{equation}
where $p_{\rm det}(\theta)\in [0,1]$ is the detection probability given a binary with parameters $\theta$. The posterior of Eq.~(\ref{poppost}) has been marginalized over the expected number of events with a scale-free prior. The hyperparameters $\lambda$ thus only capture the shape of the population distribution and not the corresponding merger rate; this is equivalent to imposing $\int p_{\rm pop}(\theta|\lambda)=1$.

For $p_{\rm pop}(\theta|\lambda)$, we use the phenomenological model referred to as \textsc{Power Law + Peak} and \textsc{Default Spin} in Ref.~\cite{2021ApJ...913L...7A}, which returns the highest Bayes factor among the options they tested. The model covers $\dim(\theta)=6$ event parameters and $\dim(\lambda) =12$ population parameters. The distribution of the primary mass $m_1$ is a superposition of a power-law component with index $\alpha$ truncated between  $m_{\rm max}$ and $m_{\rm min}$ and a Gaussian component with mean $\mu_{m}$, width $\sigma_{m}$, and mixing fraction $\lambda_{m}$. The secondary mass $m_2$ conditioned on $m_1$ follows a power-law distribution with index $\beta_{q}$. The distributions of $m_{1,2}$ are smoothed over a range $\delta_{m}$ near $m_{\rm min}$. The spin magnitudes $\chi_{1,2}$ follow a beta distribution with mean $\mu_{\chi}$ and variance $\sigma^{2}_{\chi}$. The cosines of the angles between the spins and the orbital angular momentum $\theta_{1,2}$ are distributed assuming a superposition of a uniform distribution and a truncated Gaussian with a peak at $\cos\theta_{1,2}=1$, width $\sigma_t$, and mixing fraction $\zeta$. Crucially, while we adopt the same functional form of Ref.~\cite{2021ApJ...913L...7A}, the spin tilts $\theta_{1,2}$ are here inserted at past time infinity and not at detection. The distributions of all other parameters (distance, sky location, etc.) is assumed to be independent of $\lambda$ and  equal to the prior used in the underlying single-event analyses.

The integrals at the numerator of Eq.~(\ref{poppost}) are approximated with Monte Carlo summations using samples of the posterior $p(\theta|d_i)\propto \mathcal{L}({d_i}|\theta) \pi(\theta)$ from the data release accompanying Refs.~\cite{2020MNRAS.499.3295R} (O1+O2) and \cite{2021PhRvX..11b1053A} (O3a), which in total include 44 GW events with false-alarm rate $< 1~{\rm yr}^{-1}$.
The single-event priors $\pi(\theta)$ are handled analytically with suitable reweighting factors~\cite{2021arXiv210409508C}.

For the \textsc{Power Law + Peak} and \textsc{Default Spin} model, BH masses and spins are not correlated and, consequently, the population model $p_\mathrm{pop}(\theta|\lambda)$ can be written as the product of two terms, one only including masses and one only including spins. In Ref. \cite{2021ApJ...913L...7A}, the spin part was included only in the integral of Eq.~(\ref{poppost}), and not in that of Eq.~(\ref{eqseleff}). When computing $\sigma(\lambda)$, they instead used a fixed spin distribution, thus neglecting some $\lambda$ dependencies and introducing a bias. This was motivated by the large computational cost of the search injections used to estimate $p_{\rm det}(\theta)$.

We find that a simpler $p_{\rm det}(\theta)$ prescription
(as used previously, e.g.~\cite{2019ApJ...882L..24A})
fully reproduces the results of Ref.~\cite{2021ApJ...913L...7A} while allowing for a consistent inclusion of spin effects.
In particular, we use the semianalytic approximation of Ref.~\cite{1993PhRvD..47.2198F}, assuming two data-taking periods of approximately 166 days (O1+O2~\cite{2016PhRvX...6d1015A,2019PhRvX...9c1040A}) and 150 days (O3a~\cite{2021PhRvX..11b1053A}), and a single-detector signal-to-noise ratio (SNR) threshold of 8 \cite{2016ApJS..227...14A}. SNRs are computed with representative noise curves\footnote{From \href{https://dcc.ligo.org/LIGO-P1200087-v47/public}{dcc.ligo.org/LIGO-P1200087-v47} (``early high'', for O1+O2) and \href{https://dcc.ligo.org/LIGO-T2000012/public}{dcc.ligo.org/LIGO-T2000012} (``Livingston'', for O3a).} and the \textsc{IMRPhenomPv2} waveform model~\cite{2014PhRvL.113o1101H}. The integral at the denominator of Eq.~(\ref{poppost}) is approximated with a Monte Carlo sum using samples drawn from an injected population with $p(m_1)\propto m_1^{-2.35}$, $p(m_2|m_1)\propto m_2^2$~\cite{2021PhRvX..11b1053A}, uniform spin magnitudes, spin directions with equally weighted isotropic and preferentially aligned components ($\zeta=0.5$ and $\sigma_t=0.02$), and redshifts distributed uniformly in comoving volume and source-frame time.

The prior $\pi(\lambda)$ is uniform over all 12 population parameters with limits and additional cuts as in Ref.~\cite{2021ApJ...913L...7A}. We sample $p(\lambda | d)$ using \textsc{GWPopulation}~\cite{2019PhRvD.100d3030T}, \textsc{Dynesty}~\cite{2020MNRAS.493.3132S}, and \textsc{Bilby}~\cite{2019ApJS..241...27A}.

\section{Spin propagation}

We propagate BH spin orientations across emission frequencies using the precession-averaged PN formalism first developed in Refs.~\cite{2015PhRvL.114h1103K,2015PhRvD..92f4016G}. We use an updated version of the \textsc{precession} code\footnote{See \href{https://github.com/dgerosa/precession/tree/dev}{github.com/dgerosa/precession}.} which, leveraging new analytical advancements~\cite{2017PhRvD..95j4004C,2021arXiv210610291K} and numerical recipes, is considerably more efficient than the original~\cite{2016PhRvD..93l4066G}, thus facilitating the large-scale studies presented here. For an alternative implementation of the same formalism, see Ref.~\cite{2021arXiv210711902J}.

At a finite orbital separation $r$, a BH binary is specified by the component masses $m_{1,2}$ and the dimensionless spin vectors $\boldsymbol{\chi}_{1,2}$. Of these eight parameters, the masses $m_{1,2}$, spin magnitudes $\chi_{1,2}$~\cite{2006PhRvD..74j4005B}, and effective spin $\chi_{\rm eff}$~\cite{2008PhRvD..78d4021R} are constant to at least 2PN order. Averaging over both the orbital \cite{1963PhRv..131..435P}  and the precessional phases~\cite{2015PhRvL.114h1103K,2015PhRvD..92f4016G} reduces the binary dynamics to a single ordinary differential equation ${\rm d}J/{\rm d}r$, where $J$ is the magnitude of the total angular momentum. Regularization allows evolutions to/from $r=\infty$ ($f_{\rm ref} =0$~Hz), where the spin tilts $\theta_{1,2}$ have a well-defined asymptotic limit~\cite{2015PhRvD..92f4016G}. At past time infinity, a BH binary is thus \emph{fully} specified by six quantities: $m_{1,2}$, $\chi_{1,2}$, and $\theta_{1,2}$.

First, we propagate binaries  backward in time to estimate the likelihood $\mathcal{L}({d_i}|\theta)$. For a given posterior sample at $f_{\rm ref}>0$~Hz, we obtain the corresponding orbital separation $r$ using Eq.~(4.13) of Ref.~\cite{1995PhRvD..52..821K} and the Newtonian angular momentum $L = m_1 m_2 \sqrt{r/M}$ (where $M=m_1+m_2$). The $f_{\rm ref}\to r$ conversion needs to be performed using the redshifted mass $M(1+z)$ because $f_{\rm ref}$ is a detector-frame quantity. For current LIGO events with $f_{\rm ref}=11,20$~Hz, we find $6M\lesssim r\lesssim 25M$, which is in the PN regime. We then compute $J=|  \mathbf{L} + m_1^2\boldsymbol{\chi}_1 +  m^2_2 {\boldsymbol{\chi}_2}|$, which serves as the initial condition for the precession-averaged ordinary differential equation. We only propagate posterior samples, and not prior samples. Current single-event priors are isotropic in the spin directions~\cite{2020MNRAS.499.3295R,2021PhRvX..11b1053A} and isotropicity is preserved very accurately by PN integrations~\cite{2007ApJ...661L.147B,2015PhRvD..92f4016G}, such that $\pi(\cos\theta_1,\cos\theta_2)$ is uniform both in band and at $f_{\rm ref}=0$ Hz.

Figure~\ref{backpropfig} shows backpropagated posterior distributions for the events with the largest total mass (GW190521) and the largest effective spin (GW190517\_055101). For GW190521, the values of $\theta_{1,2}$ at $f_{\rm ref}=0$ Hz shows a somewhat stronger preference for the corner of the parameter space near $(\theta_1,\theta_2)=(0,\pi)$. We also show the effective spins $\chi_{\rm eff}\in[-1,1]$ and $\chi_{\rm p}\in[0,2]$ (using the consistently averaged expression from Ref.~\cite{2021PhRvD.103f4067G}). While $\chi_{\rm eff}$ is a constant of motion and therefore does not depend on reference frequency, the backpropagated $\chi_{\rm p}$ distributions are narrower (see also \cite{2021PhRvD.104h4002B}).

Second, we  evolve binaries from the injected population forward in time to compute the detectability $p_{\rm det}(\theta)$. A binary configuration described by $m_{1,2}$, $\chi_{1,2}$, and $\theta_{1,2}$ at $r=\infty$ is first integrated down to $f_{\rm ref} = 20$~Hz with corresponding redshift $z$. At this location, we sample the precessional phase from its probability distribution function using the 2PN result of Refs.~\cite{2015PhRvL.114h1103K,2015PhRvD..92f4016G} and the orbital phase uniformly in $[0,2\pi]$. The resulting system, now well in the LIGO/Virgo band, is used to compute the SNR and thus $p_{\rm det}(\theta)$.

\begin{figure*}
\label{cornerppd}
\includegraphics[height=0.6\textwidth]{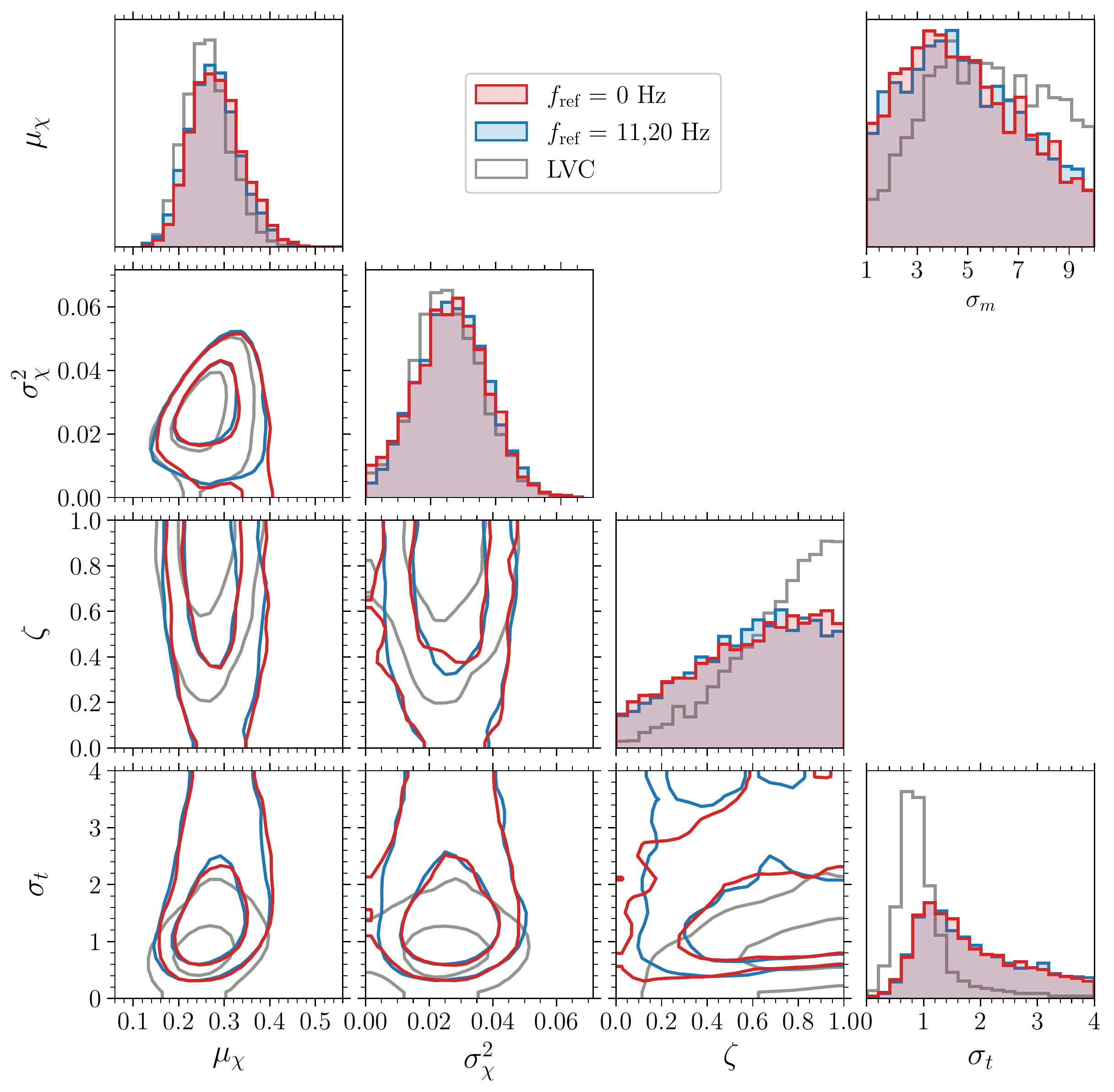}
~~~~~~~~~~
\includegraphics[height=0.6\textwidth]{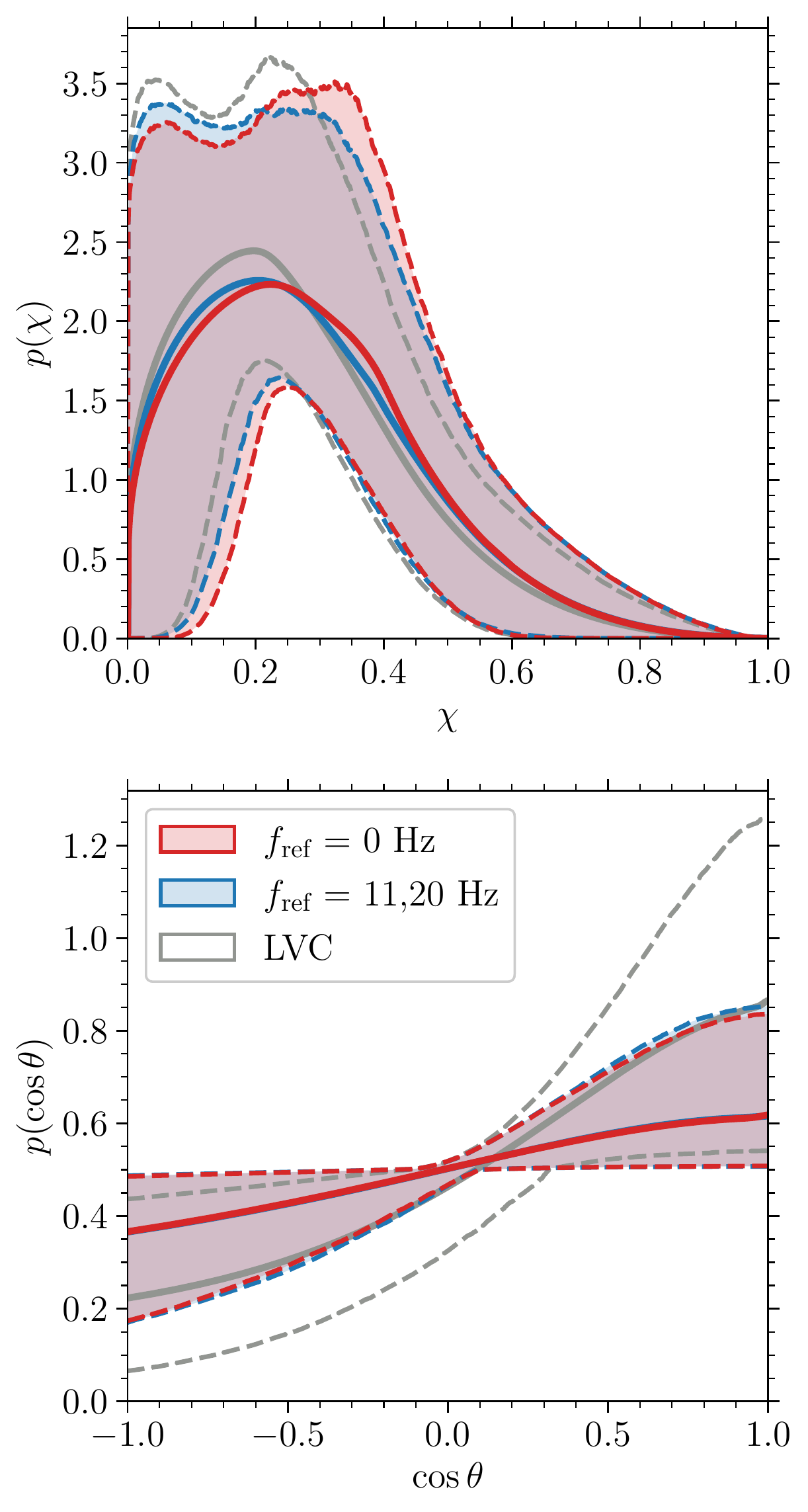}
\caption{
Left: the one- and two-dimensional marginal distributions of the spin population parameters. The red curves shows our results for population inference with spins defined at 0 Hz. The blue curves represent a control case in which spins are introduced to the selection function but the parameter estimation samples are quoted at the original reference frequency of 20 Hz (11 Hz for GW190521). The gray curves show results from the LIGO/Virgo Collaboration (LVC)~\cite{2021ApJ...913L...7A} which neglects spins when estimating selection effects. The population parameters $\mu_\chi$ and $\sigma_\chi^2$ ($\zeta$ and $\sigma_t$) describe the distributions of spin magnitudes (spin directions). The one-dimensional distribution for the width $\sigma_m$ of the Gaussian component in primary mass appears at the top-right.
Contours indicate the 50\% and 90\% credible regions.
Right: the posterior population distributions (means, bold lines) of the dimensionless spin magnitudes $\chi$ and spin tilts $\theta$. The shaded regions mark the symmetric 90\% confidence intervals bounded by the 5\% and 95\% quantiles (lower and upper dashed lines, respectively).
}
\label{cornerfig}
\end{figure*}

\section{Resulting populations}

Our results are presented in Fig.~\ref{cornerppd} where we show, for the first time, the asymptotic population of BH spins inferred from current LIGO/Virgo data ($f_{\rm ref} = 0$ Hz, red). We also run a control case with an identical setup but taking the spin tilts at detection ($f_{\rm ref} = 11,20$~Hz, blue). As in Ref.~\cite{2021ApJ...913L...7A}, in this case we use the single-event posterior samples at face value, inconsistently mixing reference frequencies. The third distribution presented in Fig.~\ref{cornerppd} (gray) is obtained from posterior samples publicly released by the LVC~\cite{2021ApJ...913L...7A}. In this case, spins are neglected when evaluating selection effects.

The corner plot on the left of Fig.~\ref{cornerppd} shows that including spins in the selection function (blue, red) has a moderately significant effect on the spin population parameters. The posterior weights are increased at small $\zeta$ (where the isotropic component of the spin-tilt distribution is favored over alignment) and large  $\sigma_t$ (increasing the width of the aligned component). In particular, we (LVC~\cite{2021ApJ...913L...7A}) find medians and 90\% intervals of $\zeta=0.63_{-0.52}^{+0.33}$ ($\zeta=0.76_{-0.48}^{+0.22}$) and $\sigma_t=1.60_{-0.91}^{+1.95}$ ($\sigma_t=0.87_{-0.45}^{+1.29}$). Therefore, we infer somewhat larger spin misalignments compared to Ref.~\cite{2021ApJ...913L...7A}. This is demonstrated in the posterior population distribution $\int p_{\rm pop}(\theta|\lambda) p(\lambda|d) {\rm d}\lambda$ on the right of Fig.~\ref{cornerppd}, where we find $p(\theta<45^\circ)= 18\%$ (red and blue) compared to 24\% (gray) for Ref.~\cite{2021ApJ...913L...7A}. Aligned binaries are easier to detect, implying that their intrinsic distribution is more heavily suppressed compared to other regions of the parameter space. This is a form of Malmquist bias equivalent to the more familiar case of the BH masses, where numerous observed events at $m\sim 30 M_\odot$ imply a lower intrinsic merger rate compared to $m\sim 5 M_\odot$, even if we have only observed a few low-mass events.

The impact on the parameters $\mu_\chi$ and $\sigma_\chi^2$ governing the spin magnitude ($\chi$) distribution, and hence the posterior population distribution for $\chi$, is less prominent. There is a minor suppression of BHs with low spin magnitudes (top-right panel in Fig.~\ref{cornerfig}), which is driven by correlations between the hyperparameters describing spin magnitudes and orientations. This is not surprising, as the best measured spin parameters are the effective quantities $\chi_{\rm eff}$ and $\chi_{\rm p}$. The mass parameters (which are not all shown in Fig.~\ref{cornerppd} for clarity) are largely unaffected by our analysis. We report a minor shift in the Gaussian component of the primary mass distribution as determined by $\sigma_m$ in Fig.~\ref{cornerfig}, which is narrower ($\sigma_m=4.62_{-3.10}^{+4.32}$) compared to the LVC result ($\sigma_m=5.69_{-3.60}^{+3.78}$).

\begin{figure}
\includegraphics[scale=0.6]{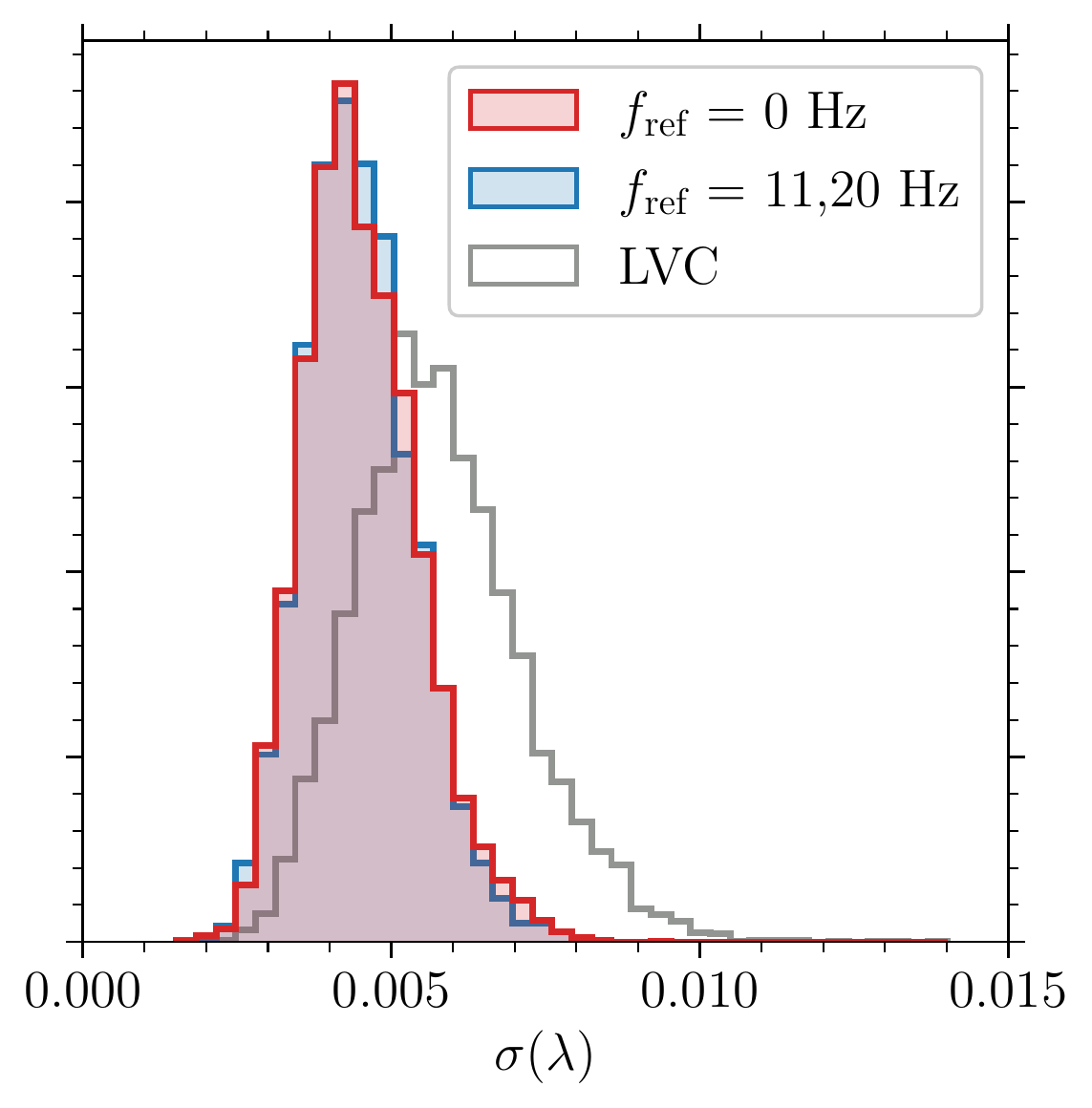}
\caption{Distributions of the selection function $\sigma(\lambda)$ across the inferred population. The gray curve displays the values provided in the data release of Ref.~\cite{2021ApJ...913L...7A}, which neglects spins when computing $\sigma$. Including spins in the selection function, in blue is the result for the control case with spins defined at the inconsistent reference frequency of 20 Hz (11 Hz for GW190521), while in red spins are defined at 0 Hz.} 
\label{sigmalambda}
\end{figure}

Figure~\ref{sigmalambda} show our Bayesian measurement of $\sigma(\lambda)$, i.e. the distribution of Eq.~(\ref{eqseleff}) across samples of Eq.~(\ref{poppost}), which sets the fraction of events from the inferred population that are observable. Including spins in the detectability returns a lower fraction $\sigma(\lambda)= 0.44^{+0.18}_{-0.13}\%$ compared to $\sigma(\lambda)= 0.57^{+0.25}_{-0.19}\%$ as obtained from the samples of Ref.~\cite{2021ApJ...913L...7A}. In other words, by neglecting spins in the inference, one is tempted to think that the intrinsic and observed population are more similar to each other than what they really are.

Once selection effects are properly accounted for, we find that considering spins at $20$ or $0$ Hz has a minor impact (compare blue and red distributions in Figs.~\ref{cornerppd} and \ref{sigmalambda}). The systematic error one incurs when putting together spin directions corresponding to different evolutionary stages of the binary inspiral is, at present, subdominant compared to statistical errors. While our findings validate the assumptions made so far in the GW literature, we stress that this is an important conceptual point which is addressed here for the first time. Furthermore, while statistical errors are bound to decrease with increasing catalog size and improved detector sensitivity, systematics will be amplified. A future publication will investigate when biases due to inconsistent parameter references are expected to become significant in population analyses.

\section{Conclusions}

We solved two conceptual issues that affected all GW population studies so far. An unbiased inference requires the BH spins to be (i) fully included when estimating selection effects and (ii) measured at past time infinity. We find that, with the same GW events and population model, the preference for spin alignment inferred in Ref.~\cite{2021ApJ...913L...7A} is reduced. Notably, this result is the opposite to the recent findings of Refs.~\cite{2021PhRvD.104h3010R,2021arXiv210902424G}, although for a very different reason. There it is claimed that the \textsc{Default Spin} model is inappropriate to describe the current set of events and underpredicts spin alignment. Future  applications of sophisticated nonparametric population modeling (e.g.~\cite{2021arXiv210905960R})
will inevitably require a consistent reference among all events.

While here we focused on BH spins, the orbital eccentricity is another key property that evolves during the binary inspiral. As GW analyses extend from quasi-circular to eccentric sources, one will need to face the issue of building a population model for the eccentricity in a coherent fashion. We anticipate this could be tackled using suitable extensions to the averaging techniques proposed here (cf. Ref.~\cite{2020PhRvD.102l3009Y}).

The spin directions at past time infinity are equal, to an excellent approximation  \cite{2021arXiv210711902J}, to those at BH formation \cite{2000ApJ...541..319K, 2008ApJ...682..474B,2013PhRvD..87j4028G}. As GW astronomy enters its large-statistics regime, traveling backward and forward in time when performing population analyses will allow for a direct, systematics-free comparison between GW data and astrophysical models of compact-object formation.

\section*{Acknowledgments}

We thank Daria Gangardt, Riccardo Buscicchio, Colm Talbot, Daniel Wysocki, and
Richard O'Shaughnessy for discussions.
M.M. and D.G. are supported by European Union's H2020 ERC Starting Grant No. 945155-GWmining, Cariplo Foundation Grant No. 2021-0555, Leverhulme Trust Grant No. RPG-2019-350, and Royal Society Grant No. RGS-R2-202004.
M.M. acknowledges networking support by the European COST Action CA16104--GWverse.
Computational work was performed on the University of Birmingham BlueBEAR cluster, the Baskerville Tier 2 HPC service funded by EPSRC Grant EP/T022221/1, and at CINECA through INFN allocations.

{\it Note added.}—Data from the second half of the third LIGO/Virgo observing run were released while this paper was being reviewed \cite{2021arXiv211103606T,2021arXiv211103634T}. The updated population analysis of Ref.~\cite{2021arXiv211103634T} includes spins in the selection effects, but does not stress its crucial importance when drawing the posterior population distributions of the spin tilts. The issue of back-propagating spins is being recognized by the wider community and was adopted in the single-event analysis of Ref.~\cite{2021arXiv211103606T}. However, when putting events together at the population level, Ref.~\cite{2021arXiv211103634T}  still uses spin directions evaluated in band.

\bibliography{draft}

\begin{thebibliography}{43}%
\makeatletter
\providecommand \@ifxundefined [1]{%
 \@ifx{#1\undefined}
}%
\providecommand \@ifnum [1]{%
 \ifnum #1\expandafter \@firstoftwo
 \else \expandafter \@secondoftwo
 \fi
}%
\providecommand \@ifx [1]{%
 \ifx #1\expandafter \@firstoftwo
 \else \expandafter \@secondoftwo
 \fi
}%
\providecommand \natexlab [1]{#1}%
\providecommand \enquote  [1]{``#1''}%
\providecommand \bibnamefont  [1]{#1}%
\providecommand \bibfnamefont [1]{#1}%
\providecommand \citenamefont [1]{#1}%
\providecommand \href@noop [0]{\@secondoftwo}%
\providecommand \href [0]{\begingroup \@sanitize@url \@href}%
\providecommand \@href[1]{\@@startlink{#1}\@@href}%
\providecommand \@@href[1]{\endgroup#1\@@endlink}%
\providecommand \@sanitize@url [0]{\catcode `\\12\catcode `\$12\catcode
  `\&12\catcode `\#12\catcode `\^12\catcode `\_12\catcode `\%12\relax}%
\providecommand \@@startlink[1]{}%
\providecommand \@@endlink[0]{}%
\providecommand \url  [0]{\begingroup\@sanitize@url \@url }%
\providecommand \@url [1]{\endgroup\@href {#1}{\urlprefix }}%
\providecommand \urlprefix  [0]{URL }%
\providecommand \Eprint [0]{\href }%
\providecommand \doibase [0]{https://doi.org/}%
\providecommand \selectlanguage [0]{\@gobble}%
\providecommand \bibinfo  [0]{\@secondoftwo}%
\providecommand \bibfield  [0]{\@secondoftwo}%
\providecommand \translation [1]{[#1]}%
\providecommand \BibitemOpen [0]{}%
\providecommand \bibitemStop [0]{}%
\providecommand \bibitemNoStop [0]{.\EOS\space}%
\providecommand \EOS [0]{\spacefactor3000\relax}%
\providecommand \BibitemShut  [1]{\csname bibitem#1\endcsname}%
\let\auto@bib@innerbib\@empty
\bibitem [{\citenamefont {{Abbott}}\ \emph
  {et~al.}(2019{\natexlab{a}})\citenamefont {{Abbott}} \emph
  {et~al.}}]{2019PhRvX...9c1040A}%
  \BibitemOpen
  \bibfield  {author} {\bibinfo {author} {\bibfnamefont {B.~P.}\ \bibnamefont
  {{Abbott}}} \emph {et~al.} (\bibinfo {collaboration} {LIGO and Virgo
  Collaboration}),\ }\href {https://doi.org/10.1103/PhysRevX.9.031040}
  {\bibfield  {journal} {\bibinfo  {journal} {\prx}\ }\textbf {\bibinfo
  {volume} {9}},\ \bibinfo {eid} {031040} (\bibinfo {year}
  {2019}{\natexlab{a}})},\ \Eprint {https://arxiv.org/abs/1811.12907}
  {arXiv:1811.12907 [astro-ph.HE]} \BibitemShut {NoStop}%
\bibitem [{\citenamefont {{Abbott}}\ \emph
  {et~al.}(2021{\natexlab{a}})\citenamefont {{Abbott}} \emph
  {et~al.}}]{2021PhRvX..11b1053A}%
  \BibitemOpen
  \bibfield  {author} {\bibinfo {author} {\bibfnamefont {R.}~\bibnamefont
  {{Abbott}}} \emph {et~al.} (\bibinfo {collaboration} {LIGO and Virgo
  Collaboration}),\ }\href {https://doi.org/10.1103/PhysRevX.11.021053}
  {\bibfield  {journal} {\bibinfo  {journal} {\prx}\ }\textbf {\bibinfo
  {volume} {11}},\ \bibinfo {eid} {021053} (\bibinfo {year}
  {2021}{\natexlab{a}})},\ \Eprint {https://arxiv.org/abs/2010.14527}
  {arXiv:2010.14527 [gr-qc]} \BibitemShut {NoStop}%
\bibitem [{\citenamefont {{Apostolatos}}\ \emph {et~al.}(1994)\citenamefont
  {{Apostolatos}}, \citenamefont {{Cutler}}, \citenamefont {{Sussman}},\ and\
  \citenamefont {{Thorne}}}]{1994PhRvD..49.6274A}%
  \BibitemOpen
  \bibfield  {author} {\bibinfo {author} {\bibfnamefont {T.~A.}\ \bibnamefont
  {{Apostolatos}}}, \bibinfo {author} {\bibfnamefont {C.}~\bibnamefont
  {{Cutler}}}, \bibinfo {author} {\bibfnamefont {G.~J.}\ \bibnamefont
  {{Sussman}}},\ and\ \bibinfo {author} {\bibfnamefont {K.~S.}\ \bibnamefont
  {{Thorne}}},\ }\href {https://doi.org/10.1103/PhysRevD.49.6274} {\bibfield
  {journal} {\bibinfo  {journal} {\prd}\ }\textbf {\bibinfo {volume} {49}},\
  \bibinfo {pages} {6274} (\bibinfo {year} {1994})}\BibitemShut {NoStop}%
\bibitem [{\citenamefont {{Peters}}\ and\ \citenamefont
  {{Mathews}}(1963)}]{1963PhRv..131..435P}%
  \BibitemOpen
  \bibfield  {author} {\bibinfo {author} {\bibfnamefont {P.~C.}\ \bibnamefont
  {{Peters}}}\ and\ \bibinfo {author} {\bibfnamefont {J.}~\bibnamefont
  {{Mathews}}},\ }\href {https://doi.org/10.1103/PhysRev.131.435} {\bibfield
  {journal} {\bibinfo  {journal} {Physical Review}\ }\textbf {\bibinfo {volume}
  {131}},\ \bibinfo {pages} {435} (\bibinfo {year} {1963})}\BibitemShut
  {NoStop}%
\bibitem [{\citenamefont {{Abbott}}\ \emph
  {et~al.}(2021{\natexlab{b}})\citenamefont {{Abbott}} \emph
  {et~al.}}]{2021ApJ...913L...7A}%
  \BibitemOpen
  \bibfield  {author} {\bibinfo {author} {\bibfnamefont {R.}~\bibnamefont
  {{Abbott}}} \emph {et~al.} (\bibinfo {collaboration} {LIGO and Virgo
  Collaboration}),\ }\href {https://doi.org/10.3847/2041-8213/abe949}
  {\bibfield  {journal} {\bibinfo  {journal} {\apjl}\ }\textbf {\bibinfo
  {volume} {913}},\ \bibinfo {eid} {L7} (\bibinfo {year}
  {2021}{\natexlab{b}})},\ \Eprint {https://arxiv.org/abs/2010.14533}
  {arXiv:2010.14533 [astro-ph.HE]} \BibitemShut {NoStop}%
\bibitem [{\citenamefont {{Varma}}\ \emph {et~al.}(2021)\citenamefont
  {{Varma}}, \citenamefont {{Isi}}, \citenamefont {{Biscoveanu}}, \citenamefont
  {{Farr}},\ and\ \citenamefont {{Vitale}}}]{2021arXiv210709692V}%
  \BibitemOpen
  \bibfield  {author} {\bibinfo {author} {\bibfnamefont {V.}~\bibnamefont
  {{Varma}}}, \bibinfo {author} {\bibfnamefont {M.}~\bibnamefont {{Isi}}},
  \bibinfo {author} {\bibfnamefont {S.}~\bibnamefont {{Biscoveanu}}}, \bibinfo
  {author} {\bibfnamefont {W.~M.}\ \bibnamefont {{Farr}}},\ and\ \bibinfo
  {author} {\bibfnamefont {S.}~\bibnamefont {{Vitale}}},\ }\href@noop {}
  {\bibfield  {journal} {\bibinfo  {journal} {{}}\ } (\bibinfo {year}
  {2021})},\ \Eprint {https://arxiv.org/abs/2107.09692} {arXiv:2107.09692
  [astro-ph.HE]} \BibitemShut {NoStop}%
\bibitem [{\citenamefont {{Mandel}}\ \emph {et~al.}(2019)\citenamefont
  {{Mandel}}, \citenamefont {{Farr}},\ and\ \citenamefont
  {{Gair}}}]{2019MNRAS.486.1086M}%
  \BibitemOpen
  \bibfield  {author} {\bibinfo {author} {\bibfnamefont {I.}~\bibnamefont
  {{Mandel}}}, \bibinfo {author} {\bibfnamefont {W.~M.}\ \bibnamefont
  {{Farr}}},\ and\ \bibinfo {author} {\bibfnamefont {J.~R.}\ \bibnamefont
  {{Gair}}},\ }\href {https://doi.org/10.1093/mnras/stz896} {\bibfield
  {journal} {\bibinfo  {journal} {\mnras}\ }\textbf {\bibinfo {volume} {486}},\
  \bibinfo {pages} {1086} (\bibinfo {year} {2019})},\ \Eprint
  {https://arxiv.org/abs/1809.02063} {arXiv:1809.02063 [physics.data-an]}
  \BibitemShut {NoStop}%
\bibitem [{\citenamefont {{Vitale}}\ \emph {et~al.}(2020)\citenamefont
  {{Vitale}}, \citenamefont {{Gerosa}}, \citenamefont {{Farr}},\ and\
  \citenamefont {{Taylor}}}]{2020arXiv200705579V}%
  \BibitemOpen
  \bibfield  {author} {\bibinfo {author} {\bibfnamefont {S.}~\bibnamefont
  {{Vitale}}}, \bibinfo {author} {\bibfnamefont {D.}~\bibnamefont {{Gerosa}}},
  \bibinfo {author} {\bibfnamefont {W.~M.}\ \bibnamefont {{Farr}}},\ and\
  \bibinfo {author} {\bibfnamefont {S.~R.}\ \bibnamefont {{Taylor}}},\
  }\href@noop {} {\bibfield  {journal} {\bibinfo  {journal} {{}}\ } (\bibinfo
  {year} {2020})},\ \Eprint {https://arxiv.org/abs/2007.05579}
  {arXiv:2007.05579 [astro-ph.IM]} \BibitemShut {NoStop}%
\bibitem [{\citenamefont {{Damour}}(2001)}]{2001PhRvD..64l4013D}%
  \BibitemOpen
  \bibfield  {author} {\bibinfo {author} {\bibfnamefont {T.}~\bibnamefont
  {{Damour}}},\ }\href {https://doi.org/10.1103/PhysRevD.64.124013} {\bibfield
  {journal} {\bibinfo  {journal} {\prd}\ }\textbf {\bibinfo {volume} {64}},\
  \bibinfo {pages} {124013} (\bibinfo {year} {2001})},\ \Eprint
  {https://arxiv.org/abs/gr-qc/0103018} {arXiv:gr-qc/0103018 [gr-qc]}
  \BibitemShut {NoStop}%
\bibitem [{\citenamefont {{Ng}}\ \emph {et~al.}(2018)\citenamefont {{Ng}},
  \citenamefont {{Vitale}}, \citenamefont {{Zimmerman}}, \citenamefont
  {{Chatziioannou}}, \citenamefont {{Gerosa}},\ and\ \citenamefont
  {{Haster}}}]{2018PhRvD..98h3007N}%
  \BibitemOpen
  \bibfield  {author} {\bibinfo {author} {\bibfnamefont {K.~K.~Y.}\
  \bibnamefont {{Ng}}}, \bibinfo {author} {\bibfnamefont {S.}~\bibnamefont
  {{Vitale}}}, \bibinfo {author} {\bibfnamefont {A.}~\bibnamefont
  {{Zimmerman}}}, \bibinfo {author} {\bibfnamefont {K.}~\bibnamefont
  {{Chatziioannou}}}, \bibinfo {author} {\bibfnamefont {D.}~\bibnamefont
  {{Gerosa}}},\ and\ \bibinfo {author} {\bibfnamefont {C.-J.}\ \bibnamefont
  {{Haster}}},\ }\href {https://doi.org/10.1103/PhysRevD.98.083007} {\bibfield
  {journal} {\bibinfo  {journal} {\prd}\ }\textbf {\bibinfo {volume} {98}},\
  \bibinfo {eid} {083007} (\bibinfo {year} {2018})},\ \Eprint
  {https://arxiv.org/abs/1805.03046} {arXiv:1805.03046 [gr-qc]} \BibitemShut
  {NoStop}%
\bibitem [{\citenamefont {{Gerosa}}\ \emph {et~al.}(2015)\citenamefont
  {{Gerosa}}, \citenamefont {{Kesden}}, \citenamefont {{Sperhake}},
  \citenamefont {{Berti}},\ and\ \citenamefont
  {{O'Shaughnessy}}}]{2015PhRvD..92f4016G}%
  \BibitemOpen
  \bibfield  {author} {\bibinfo {author} {\bibfnamefont {D.}~\bibnamefont
  {{Gerosa}}}, \bibinfo {author} {\bibfnamefont {M.}~\bibnamefont {{Kesden}}},
  \bibinfo {author} {\bibfnamefont {U.}~\bibnamefont {{Sperhake}}}, \bibinfo
  {author} {\bibfnamefont {E.}~\bibnamefont {{Berti}}},\ and\ \bibinfo {author}
  {\bibfnamefont {R.}~\bibnamefont {{O'Shaughnessy}}},\ }\href
  {https://doi.org/10.1103/PhysRevD.92.064016} {\bibfield  {journal} {\bibinfo
  {journal} {\prd}\ }\textbf {\bibinfo {volume} {92}},\ \bibinfo {eid} {064016}
  (\bibinfo {year} {2015})},\ \Eprint {https://arxiv.org/abs/1506.03492}
  {arXiv:1506.03492 [gr-qc]} \BibitemShut {NoStop}%
\bibitem [{\citenamefont {{Reali}}\ \emph {et~al.}(2020)\citenamefont
  {{Reali}}, \citenamefont {{Mould}}, \citenamefont {{Gerosa}},\ and\
  \citenamefont {{Varma}}}]{2020CQGra..37v5005R}%
  \BibitemOpen
  \bibfield  {author} {\bibinfo {author} {\bibfnamefont {L.}~\bibnamefont
  {{Reali}}}, \bibinfo {author} {\bibfnamefont {M.}~\bibnamefont {{Mould}}},
  \bibinfo {author} {\bibfnamefont {D.}~\bibnamefont {{Gerosa}}},\ and\
  \bibinfo {author} {\bibfnamefont {V.}~\bibnamefont {{Varma}}},\ }\href
  {https://doi.org/10.1088/1361-6382/abb639} {\bibfield  {journal} {\bibinfo
  {journal} {\cqg}\ }\textbf {\bibinfo {volume} {37}},\ \bibinfo {eid} {225005}
  (\bibinfo {year} {2020})},\ \Eprint {https://arxiv.org/abs/2005.01747}
  {arXiv:2005.01747 [gr-qc]} \BibitemShut {NoStop}%
\bibitem [{\citenamefont {{Loredo}}(2004)}]{2004AIPC..735..195L}%
  \BibitemOpen
  \bibfield  {author} {\bibinfo {author} {\bibfnamefont {T.~J.}\ \bibnamefont
  {{Loredo}}},\ }\href {https://doi.org/10.1063/1.1835214} {\bibfield
  {journal} {\bibinfo  {journal} {AIP Conf. Proc.}\ }\textbf {\bibinfo {volume}
  {735}},\ \bibinfo {pages} {195} (\bibinfo {year} {2004})},\ \Eprint
  {https://arxiv.org/abs/astro-ph/0409387} {arXiv:astro-ph/0409387 [astro-ph]}
  \BibitemShut {NoStop}%
\bibitem [{\citenamefont {{Romero-Shaw}}\ \emph {et~al.}(2020)\citenamefont
  {{Romero-Shaw}}, \citenamefont {{Talbot}}, \citenamefont {{Biscoveanu}} \emph
  {et~al.}}]{2020MNRAS.499.3295R}%
  \BibitemOpen
  \bibfield  {author} {\bibinfo {author} {\bibfnamefont {I.~M.}\ \bibnamefont
  {{Romero-Shaw}}}, \bibinfo {author} {\bibfnamefont {C.}~\bibnamefont
  {{Talbot}}}, \bibinfo {author} {\bibfnamefont {S.}~\bibnamefont
  {{Biscoveanu}}}, \emph {et~al.},\ }\href
  {https://doi.org/10.1093/mnras/staa2850} {\bibfield  {journal} {\bibinfo
  {journal} {\mnras}\ }\textbf {\bibinfo {volume} {499}},\ \bibinfo {pages}
  {3295} (\bibinfo {year} {2020})},\ \Eprint {https://arxiv.org/abs/2006.00714}
  {arXiv:2006.00714 [astro-ph.IM]} \BibitemShut {NoStop}%
\bibitem [{\citenamefont {{Callister}}(2021)}]{2021arXiv210409508C}%
  \BibitemOpen
  \bibfield  {author} {\bibinfo {author} {\bibfnamefont {T.~A.}\ \bibnamefont
  {{Callister}}},\ }\href@noop {} {\bibfield  {journal} {\bibinfo  {journal}
  {{}}\ } (\bibinfo {year} {2021})},\ \Eprint
  {https://arxiv.org/abs/2104.09508} {arXiv:2104.09508 [gr-qc]} \BibitemShut
  {NoStop}%
\bibitem [{\citenamefont {{Abbott}}\ \emph
  {et~al.}(2019{\natexlab{b}})\citenamefont {{Abbott}} \emph
  {et~al.}}]{2019ApJ...882L..24A}%
  \BibitemOpen
  \bibfield  {author} {\bibinfo {author} {\bibfnamefont {B.~P.}\ \bibnamefont
  {{Abbott}}} \emph {et~al.} (\bibinfo {collaboration} {LIGO and Virgo
  Collaboration}),\ }\href {https://doi.org/10.3847/2041-8213/ab3800}
  {\bibfield  {journal} {\bibinfo  {journal} {\apjl}\ }\textbf {\bibinfo
  {volume} {882}},\ \bibinfo {eid} {L24} (\bibinfo {year}
  {2019}{\natexlab{b}})},\ \Eprint {https://arxiv.org/abs/1811.12940}
  {arXiv:1811.12940 [astro-ph.HE]} \BibitemShut {NoStop}%
\bibitem [{\citenamefont {{Finn}}\ and\ \citenamefont
  {{Chernoff}}(1993)}]{1993PhRvD..47.2198F}%
  \BibitemOpen
  \bibfield  {author} {\bibinfo {author} {\bibfnamefont {L.~S.}\ \bibnamefont
  {{Finn}}}\ and\ \bibinfo {author} {\bibfnamefont {D.~F.}\ \bibnamefont
  {{Chernoff}}},\ }\href {https://doi.org/10.1103/PhysRevD.47.2198} {\bibfield
  {journal} {\bibinfo  {journal} {\prd}\ }\textbf {\bibinfo {volume} {47}},\
  \bibinfo {pages} {2198} (\bibinfo {year} {1993})},\ \Eprint
  {https://arxiv.org/abs/gr-qc/9301003} {arXiv:gr-qc/9301003 [gr-qc]}
  \BibitemShut {NoStop}%
\bibitem [{\citenamefont {{Abbott}}\ \emph
  {et~al.}(2016{\natexlab{a}})\citenamefont {{Abbott}} \emph
  {et~al.}}]{2016PhRvX...6d1015A}%
  \BibitemOpen
  \bibfield  {author} {\bibinfo {author} {\bibfnamefont {B.~P.}\ \bibnamefont
  {{Abbott}}} \emph {et~al.} (\bibinfo {collaboration} {LIGO and Virgo
  Collaboration}),\ }\href {https://doi.org/10.1103/PhysRevX.6.041015}
  {\bibfield  {journal} {\bibinfo  {journal} {\prx}\ }\textbf {\bibinfo
  {volume} {6}},\ \bibinfo {eid} {041015} (\bibinfo {year}
  {2016}{\natexlab{a}})},\ \Eprint {https://arxiv.org/abs/1606.04856}
  {arXiv:1606.04856 [gr-qc]} \BibitemShut {NoStop}%
\bibitem [{\citenamefont {{Abbott}}\ \emph
  {et~al.}(2016{\natexlab{b}})\citenamefont {{Abbott}} \emph
  {et~al.}}]{2016ApJS..227...14A}%
  \BibitemOpen
  \bibfield  {author} {\bibinfo {author} {\bibfnamefont {B.~P.}\ \bibnamefont
  {{Abbott}}} \emph {et~al.} (\bibinfo {collaboration} {LIGO and Virgo
  Collaboration}),\ }\href {https://doi.org/10.3847/0067-0049/227/2/14}
  {\bibfield  {journal} {\bibinfo  {journal} {\apjs}\ }\textbf {\bibinfo
  {volume} {227}},\ \bibinfo {eid} {14} (\bibinfo {year}
  {2016}{\natexlab{b}})},\ \Eprint {https://arxiv.org/abs/1606.03939}
  {arXiv:1606.03939 [astro-ph.HE]} \BibitemShut {NoStop}%
\bibitem [{\citenamefont {{Hannam}}\ \emph {et~al.}(2014)\citenamefont
  {{Hannam}}, \citenamefont {{Schmidt}}, \citenamefont {{Boh{\'e}}},
  \citenamefont {{Haegel}}, \citenamefont {{Husa}}, \citenamefont {{Ohme}},
  \citenamefont {{Pratten}},\ and\ \citenamefont
  {{P{\"u}rrer}}}]{2014PhRvL.113o1101H}%
  \BibitemOpen
  \bibfield  {author} {\bibinfo {author} {\bibfnamefont {M.}~\bibnamefont
  {{Hannam}}}, \bibinfo {author} {\bibfnamefont {P.}~\bibnamefont {{Schmidt}}},
  \bibinfo {author} {\bibfnamefont {A.}~\bibnamefont {{Boh{\'e}}}}, \bibinfo
  {author} {\bibfnamefont {L.}~\bibnamefont {{Haegel}}}, \bibinfo {author}
  {\bibfnamefont {S.}~\bibnamefont {{Husa}}}, \bibinfo {author} {\bibfnamefont
  {F.}~\bibnamefont {{Ohme}}}, \bibinfo {author} {\bibfnamefont
  {G.}~\bibnamefont {{Pratten}}},\ and\ \bibinfo {author} {\bibfnamefont
  {M.}~\bibnamefont {{P{\"u}rrer}}},\ }\href
  {https://doi.org/10.1103/PhysRevLett.113.151101} {\bibfield  {journal}
  {\bibinfo  {journal} {\prl}\ }\textbf {\bibinfo {volume} {113}},\ \bibinfo
  {eid} {151101} (\bibinfo {year} {2014})},\ \Eprint
  {https://arxiv.org/abs/1308.3271} {arXiv:1308.3271 [gr-qc]} \BibitemShut
  {NoStop}%
\bibitem [{\citenamefont {{Talbot}}\ \emph {et~al.}(2019)\citenamefont
  {{Talbot}}, \citenamefont {{Smith}}, \citenamefont {{Thrane}},\ and\
  \citenamefont {{Poole}}}]{2019PhRvD.100d3030T}%
  \BibitemOpen
  \bibfield  {author} {\bibinfo {author} {\bibfnamefont {C.}~\bibnamefont
  {{Talbot}}}, \bibinfo {author} {\bibfnamefont {R.}~\bibnamefont {{Smith}}},
  \bibinfo {author} {\bibfnamefont {E.}~\bibnamefont {{Thrane}}},\ and\
  \bibinfo {author} {\bibfnamefont {G.~B.}\ \bibnamefont {{Poole}}},\ }\href
  {https://doi.org/10.1103/PhysRevD.100.043030} {\bibfield  {journal} {\bibinfo
   {journal} {\prd}\ }\textbf {\bibinfo {volume} {100}},\ \bibinfo {eid}
  {043030} (\bibinfo {year} {2019})},\ \Eprint
  {https://arxiv.org/abs/1904.02863} {arXiv:1904.02863 [astro-ph.IM]}
  \BibitemShut {NoStop}%
\bibitem [{\citenamefont {{Speagle}}(2020)}]{2020MNRAS.493.3132S}%
  \BibitemOpen
  \bibfield  {author} {\bibinfo {author} {\bibfnamefont {J.~S.}\ \bibnamefont
  {{Speagle}}},\ }\href {https://doi.org/10.1093/mnras/staa278} {\bibfield
  {journal} {\bibinfo  {journal} {\mnras}\ }\textbf {\bibinfo {volume} {493}},\
  \bibinfo {pages} {3132} (\bibinfo {year} {2020})},\ \Eprint
  {https://arxiv.org/abs/1904.02180} {arXiv:1904.02180 [astro-ph.IM]}
  \BibitemShut {NoStop}%
\bibitem [{\citenamefont {{Ashton}}\ \emph {et~al.}(2019)\citenamefont
  {{Ashton}}, \citenamefont {{H{\"u}bner}}, \citenamefont {{Lasky}},
  \citenamefont {{Talbot}}, \citenamefont {{Ackley}}, \citenamefont
  {{Biscoveanu}}, \citenamefont {{Chu}}, \citenamefont {{Divakarla}},
  \citenamefont {{Easter}}, \citenamefont {{Goncharov}}, \citenamefont
  {{Hernandez Vivanco}}, \citenamefont {{Harms}}, \citenamefont {{Lower}},
  \citenamefont {{Meadors}}, \citenamefont {{Melchor}}, \citenamefont
  {{Payne}}, \citenamefont {{Pitkin}}, \citenamefont {{Powell}}, \citenamefont
  {{Sarin}}, \citenamefont {{Smith}},\ and\ \citenamefont
  {{Thrane}}}]{2019ApJS..241...27A}%
  \BibitemOpen
  \bibfield  {author} {\bibinfo {author} {\bibfnamefont {G.}~\bibnamefont
  {{Ashton}}}, \bibinfo {author} {\bibfnamefont {M.}~\bibnamefont
  {{H{\"u}bner}}}, \bibinfo {author} {\bibfnamefont {P.~D.}\ \bibnamefont
  {{Lasky}}}, \bibinfo {author} {\bibfnamefont {C.}~\bibnamefont {{Talbot}}},
  \bibinfo {author} {\bibfnamefont {K.}~\bibnamefont {{Ackley}}}, \bibinfo
  {author} {\bibfnamefont {S.}~\bibnamefont {{Biscoveanu}}}, \bibinfo {author}
  {\bibfnamefont {Q.}~\bibnamefont {{Chu}}}, \bibinfo {author} {\bibfnamefont
  {A.}~\bibnamefont {{Divakarla}}}, \bibinfo {author} {\bibfnamefont {P.~J.}\
  \bibnamefont {{Easter}}}, \bibinfo {author} {\bibfnamefont {B.}~\bibnamefont
  {{Goncharov}}}, \bibinfo {author} {\bibfnamefont {F.}~\bibnamefont
  {{Hernandez Vivanco}}}, \bibinfo {author} {\bibfnamefont {J.}~\bibnamefont
  {{Harms}}}, \bibinfo {author} {\bibfnamefont {M.~E.}\ \bibnamefont
  {{Lower}}}, \bibinfo {author} {\bibfnamefont {G.~D.}\ \bibnamefont
  {{Meadors}}}, \bibinfo {author} {\bibfnamefont {D.}~\bibnamefont
  {{Melchor}}}, \bibinfo {author} {\bibfnamefont {E.}~\bibnamefont {{Payne}}},
  \bibinfo {author} {\bibfnamefont {M.~D.}\ \bibnamefont {{Pitkin}}}, \bibinfo
  {author} {\bibfnamefont {J.}~\bibnamefont {{Powell}}}, \bibinfo {author}
  {\bibfnamefont {N.}~\bibnamefont {{Sarin}}}, \bibinfo {author} {\bibfnamefont
  {R.~J.~E.}\ \bibnamefont {{Smith}}},\ and\ \bibinfo {author} {\bibfnamefont
  {E.}~\bibnamefont {{Thrane}}},\ }\href
  {https://doi.org/10.3847/1538-4365/ab06fc} {\bibfield  {journal} {\bibinfo
  {journal} {\apjs}\ }\textbf {\bibinfo {volume} {241}},\ \bibinfo {eid} {27}
  (\bibinfo {year} {2019})},\ \Eprint {https://arxiv.org/abs/1811.02042}
  {arXiv:1811.02042 [astro-ph.IM]} \BibitemShut {NoStop}%
\bibitem [{\citenamefont {{Kesden}}\ \emph {et~al.}(2015)\citenamefont
  {{Kesden}}, \citenamefont {{Gerosa}}, \citenamefont {{O'Shaughnessy}},
  \citenamefont {{Berti}},\ and\ \citenamefont
  {{Sperhake}}}]{2015PhRvL.114h1103K}%
  \BibitemOpen
  \bibfield  {author} {\bibinfo {author} {\bibfnamefont {M.}~\bibnamefont
  {{Kesden}}}, \bibinfo {author} {\bibfnamefont {D.}~\bibnamefont {{Gerosa}}},
  \bibinfo {author} {\bibfnamefont {R.}~\bibnamefont {{O'Shaughnessy}}},
  \bibinfo {author} {\bibfnamefont {E.}~\bibnamefont {{Berti}}},\ and\ \bibinfo
  {author} {\bibfnamefont {U.}~\bibnamefont {{Sperhake}}},\ }\href
  {https://doi.org/10.1103/PhysRevLett.114.081103} {\bibfield  {journal}
  {\bibinfo  {journal} {\prl}\ }\textbf {\bibinfo {volume} {114}},\ \bibinfo
  {eid} {081103} (\bibinfo {year} {2015})},\ \Eprint
  {https://arxiv.org/abs/1411.0674} {arXiv:1411.0674 [gr-qc]} \BibitemShut
  {NoStop}%
\bibitem [{\citenamefont {{Chatziioannou}}\ \emph {et~al.}(2017)\citenamefont
  {{Chatziioannou}}, \citenamefont {{Klein}}, \citenamefont {{Yunes}},\ and\
  \citenamefont {{Cornish}}}]{2017PhRvD..95j4004C}%
  \BibitemOpen
  \bibfield  {author} {\bibinfo {author} {\bibfnamefont {K.}~\bibnamefont
  {{Chatziioannou}}}, \bibinfo {author} {\bibfnamefont {A.}~\bibnamefont
  {{Klein}}}, \bibinfo {author} {\bibfnamefont {N.}~\bibnamefont {{Yunes}}},\
  and\ \bibinfo {author} {\bibfnamefont {N.}~\bibnamefont {{Cornish}}},\ }\href
  {https://doi.org/10.1103/PhysRevD.95.104004} {\bibfield  {journal} {\bibinfo
  {journal} {\prd}\ }\textbf {\bibinfo {volume} {95}},\ \bibinfo {eid} {104004}
  (\bibinfo {year} {2017})},\ \Eprint {https://arxiv.org/abs/1703.03967}
  {arXiv:1703.03967 [gr-qc]} \BibitemShut {NoStop}%
\bibitem [{\citenamefont {{Klein}}(2021)}]{2021arXiv210610291K}%
  \BibitemOpen
  \bibfield  {author} {\bibinfo {author} {\bibfnamefont {A.}~\bibnamefont
  {{Klein}}},\ }\href@noop {} {\bibfield  {journal} {\bibinfo  {journal} {{}}\
  } (\bibinfo {year} {2021})},\ \Eprint {https://arxiv.org/abs/2106.10291}
  {arXiv:2106.10291 [gr-qc]} \BibitemShut {NoStop}%
\bibitem [{\citenamefont {{Gerosa}}\ and\ \citenamefont
  {{Kesden}}(2016)}]{2016PhRvD..93l4066G}%
  \BibitemOpen
  \bibfield  {author} {\bibinfo {author} {\bibfnamefont {D.}~\bibnamefont
  {{Gerosa}}}\ and\ \bibinfo {author} {\bibfnamefont {M.}~\bibnamefont
  {{Kesden}}},\ }\href {https://doi.org/10.1103/PhysRevD.93.124066} {\bibfield
  {journal} {\bibinfo  {journal} {\prd}\ }\textbf {\bibinfo {volume} {93}},\
  \bibinfo {eid} {124066} (\bibinfo {year} {2016})},\ \Eprint
  {https://arxiv.org/abs/1605.01067} {arXiv:1605.01067 [astro-ph.HE]}
  \BibitemShut {NoStop}%
\bibitem [{\citenamefont {{Johnson-McDaniel}}\ \emph
  {et~al.}(2021)\citenamefont {{Johnson-McDaniel}}, \citenamefont
  {{Kulkarni}},\ and\ \citenamefont {{Gupta}}}]{2021arXiv210711902J}%
  \BibitemOpen
  \bibfield  {author} {\bibinfo {author} {\bibfnamefont {N.~K.}\ \bibnamefont
  {{Johnson-McDaniel}}}, \bibinfo {author} {\bibfnamefont {S.}~\bibnamefont
  {{Kulkarni}}},\ and\ \bibinfo {author} {\bibfnamefont {A.}~\bibnamefont
  {{Gupta}}},\ }\href@noop {} {\bibfield  {journal} {\bibinfo  {journal} {{}}\
  } (\bibinfo {year} {2021})},\ \Eprint {https://arxiv.org/abs/2107.11902}
  {arXiv:2107.11902 [astro-ph.HE]} \BibitemShut {NoStop}%
\bibitem [{\citenamefont {{Buonanno}}\ \emph {et~al.}(2006)\citenamefont
  {{Buonanno}}, \citenamefont {{Chen}},\ and\ \citenamefont
  {{Damour}}}]{2006PhRvD..74j4005B}%
  \BibitemOpen
  \bibfield  {author} {\bibinfo {author} {\bibfnamefont {A.}~\bibnamefont
  {{Buonanno}}}, \bibinfo {author} {\bibfnamefont {Y.}~\bibnamefont {{Chen}}},\
  and\ \bibinfo {author} {\bibfnamefont {T.}~\bibnamefont {{Damour}}},\ }\href
  {https://doi.org/10.1103/PhysRevD.74.104005} {\bibfield  {journal} {\bibinfo
  {journal} {\prd}\ }\textbf {\bibinfo {volume} {74}},\ \bibinfo {eid} {104005}
  (\bibinfo {year} {2006})},\ \Eprint {https://arxiv.org/abs/gr-qc/0508067}
  {arXiv:gr-qc/0508067 [gr-qc]} \BibitemShut {NoStop}%
\bibitem [{\citenamefont {{Racine}}(2008)}]{2008PhRvD..78d4021R}%
  \BibitemOpen
  \bibfield  {author} {\bibinfo {author} {\bibfnamefont {{\'E}.}~\bibnamefont
  {{Racine}}},\ }\href {https://doi.org/10.1103/PhysRevD.78.044021} {\bibfield
  {journal} {\bibinfo  {journal} {\prd}\ }\textbf {\bibinfo {volume} {78}},\
  \bibinfo {eid} {044021} (\bibinfo {year} {2008})},\ \Eprint
  {https://arxiv.org/abs/0803.1820} {arXiv:0803.1820 [gr-qc]} \BibitemShut
  {NoStop}%
\bibitem [{\citenamefont {{Kidder}}(1995)}]{1995PhRvD..52..821K}%
  \BibitemOpen
  \bibfield  {author} {\bibinfo {author} {\bibfnamefont {L.~E.}\ \bibnamefont
  {{Kidder}}},\ }\href {https://doi.org/10.1103/PhysRevD.52.821} {\bibfield
  {journal} {\bibinfo  {journal} {\prd}\ }\textbf {\bibinfo {volume} {52}},\
  \bibinfo {pages} {821} (\bibinfo {year} {1995})},\ \Eprint
  {https://arxiv.org/abs/gr-qc/9506022} {arXiv:gr-qc/9506022 [gr-qc]}
  \BibitemShut {NoStop}%
\bibitem [{\citenamefont {{Bogdanovi{\'c}}}\ \emph {et~al.}(2007)\citenamefont
  {{Bogdanovi{\'c}}}, \citenamefont {{Reynolds}},\ and\ \citenamefont
  {{Miller}}}]{2007ApJ...661L.147B}%
  \BibitemOpen
  \bibfield  {author} {\bibinfo {author} {\bibfnamefont {T.}~\bibnamefont
  {{Bogdanovi{\'c}}}}, \bibinfo {author} {\bibfnamefont {C.~S.}\ \bibnamefont
  {{Reynolds}}},\ and\ \bibinfo {author} {\bibfnamefont {M.~C.}\ \bibnamefont
  {{Miller}}},\ }\href {https://doi.org/10.1086/518769} {\bibfield  {journal}
  {\bibinfo  {journal} {\apjl}\ }\textbf {\bibinfo {volume} {661}},\ \bibinfo
  {pages} {L147} (\bibinfo {year} {2007})},\ \Eprint
  {https://arxiv.org/abs/astro-ph/0703054} {arXiv:astro-ph/0703054 [astro-ph]}
  \BibitemShut {NoStop}%
\bibitem [{\citenamefont {{Gerosa}}\ \emph {et~al.}(2021)\citenamefont
  {{Gerosa}}, \citenamefont {{Mould}}, \citenamefont {{Gangardt}},
  \citenamefont {{Schmidt}}, \citenamefont {{Pratten}},\ and\ \citenamefont
  {{Thomas}}}]{2021PhRvD.103f4067G}%
  \BibitemOpen
  \bibfield  {author} {\bibinfo {author} {\bibfnamefont {D.}~\bibnamefont
  {{Gerosa}}}, \bibinfo {author} {\bibfnamefont {M.}~\bibnamefont {{Mould}}},
  \bibinfo {author} {\bibfnamefont {D.}~\bibnamefont {{Gangardt}}}, \bibinfo
  {author} {\bibfnamefont {P.}~\bibnamefont {{Schmidt}}}, \bibinfo {author}
  {\bibfnamefont {G.}~\bibnamefont {{Pratten}}},\ and\ \bibinfo {author}
  {\bibfnamefont {L.~M.}\ \bibnamefont {{Thomas}}},\ }\href
  {https://doi.org/10.1103/PhysRevD.103.064067} {\bibfield  {journal} {\bibinfo
   {journal} {\prd}\ }\textbf {\bibinfo {volume} {103}},\ \bibinfo {eid}
  {064067} (\bibinfo {year} {2021})},\ \Eprint
  {https://arxiv.org/abs/2011.11948} {arXiv:2011.11948 [gr-qc]} \BibitemShut
  {NoStop}%
\bibitem [{\citenamefont {{Baibhav}}\ \emph {et~al.}(2021)\citenamefont
  {{Baibhav}}, \citenamefont {{Berti}}, \citenamefont {{Gerosa}}, \citenamefont
  {{Mould}},\ and\ \citenamefont {{Wong}}}]{2021PhRvD.104h4002B}%
  \BibitemOpen
  \bibfield  {author} {\bibinfo {author} {\bibfnamefont {V.}~\bibnamefont
  {{Baibhav}}}, \bibinfo {author} {\bibfnamefont {E.}~\bibnamefont {{Berti}}},
  \bibinfo {author} {\bibfnamefont {D.}~\bibnamefont {{Gerosa}}}, \bibinfo
  {author} {\bibfnamefont {M.}~\bibnamefont {{Mould}}},\ and\ \bibinfo {author}
  {\bibfnamefont {K.~W.~K.}\ \bibnamefont {{Wong}}},\ }\href
  {https://doi.org/10.1103/PhysRevD.104.084002} {\bibfield  {journal} {\bibinfo
   {journal} {\prd}\ }\textbf {\bibinfo {volume} {104}},\ \bibinfo {eid}
  {084002} (\bibinfo {year} {2021})},\ \Eprint
  {https://arxiv.org/abs/2105.12140} {arXiv:2105.12140 [gr-qc]} \BibitemShut
  {NoStop}%
\bibitem [{\citenamefont {{Roulet}}\ \emph {et~al.}(2021)\citenamefont
  {{Roulet}}, \citenamefont {{Chia}}, \citenamefont {{Olsen}}, \citenamefont
  {{Dai}}, \citenamefont {{Venumadhav}}, \citenamefont {{Zackay}},\ and\
  \citenamefont {{Zaldarriaga}}}]{2021PhRvD.104h3010R}%
  \BibitemOpen
  \bibfield  {author} {\bibinfo {author} {\bibfnamefont {J.}~\bibnamefont
  {{Roulet}}}, \bibinfo {author} {\bibfnamefont {H.~S.}\ \bibnamefont
  {{Chia}}}, \bibinfo {author} {\bibfnamefont {S.}~\bibnamefont {{Olsen}}},
  \bibinfo {author} {\bibfnamefont {L.}~\bibnamefont {{Dai}}}, \bibinfo
  {author} {\bibfnamefont {T.}~\bibnamefont {{Venumadhav}}}, \bibinfo {author}
  {\bibfnamefont {B.}~\bibnamefont {{Zackay}}},\ and\ \bibinfo {author}
  {\bibfnamefont {M.}~\bibnamefont {{Zaldarriaga}}},\ }\href
  {https://doi.org/10.1103/PhysRevD.104.083010} {\bibfield  {journal} {\bibinfo
   {journal} {\prd}\ }\textbf {\bibinfo {volume} {104}},\ \bibinfo {eid}
  {083010} (\bibinfo {year} {2021})},\ \Eprint
  {https://arxiv.org/abs/2105.10580} {arXiv:2105.10580 [astro-ph.HE]}
  \BibitemShut {NoStop}%
\bibitem [{\citenamefont {{Galaudage}}\ \emph {et~al.}(2021)\citenamefont
  {{Galaudage}}, \citenamefont {{Talbot}}, \citenamefont {{Nagar}},
  \citenamefont {{Jain}}, \citenamefont {{Thrane}},\ and\ \citenamefont
  {{Mandel}}}]{2021arXiv210902424G}%
  \BibitemOpen
  \bibfield  {author} {\bibinfo {author} {\bibfnamefont {S.}~\bibnamefont
  {{Galaudage}}}, \bibinfo {author} {\bibfnamefont {C.}~\bibnamefont
  {{Talbot}}}, \bibinfo {author} {\bibfnamefont {T.}~\bibnamefont {{Nagar}}},
  \bibinfo {author} {\bibfnamefont {D.}~\bibnamefont {{Jain}}}, \bibinfo
  {author} {\bibfnamefont {E.}~\bibnamefont {{Thrane}}},\ and\ \bibinfo
  {author} {\bibfnamefont {I.}~\bibnamefont {{Mandel}}},\ }\href@noop {}
  {\bibfield  {journal} {\bibinfo  {journal} {{}}\ } (\bibinfo {year}
  {2021})},\ \Eprint {https://arxiv.org/abs/2109.02424} {arXiv:2109.02424
  [gr-qc]} \BibitemShut {NoStop}%
\bibitem [{\citenamefont {{Rinaldi}}\ and\ \citenamefont {{Del
  Pozzo}}(2021)}]{2021arXiv210905960R}%
  \BibitemOpen
  \bibfield  {author} {\bibinfo {author} {\bibfnamefont {S.}~\bibnamefont
  {{Rinaldi}}}\ and\ \bibinfo {author} {\bibfnamefont {W.}~\bibnamefont {{Del
  Pozzo}}},\ }\href@noop {} {\bibfield  {journal} {\bibinfo  {journal} {{}}\ }
  (\bibinfo {year} {2021})},\ \Eprint {https://arxiv.org/abs/2109.05960}
  {arXiv:2109.05960 [astro-ph.IM]} \BibitemShut {NoStop}%
\bibitem [{\citenamefont {{Yu}}\ \emph {et~al.}(2020)\citenamefont {{Yu}},
  \citenamefont {{Ma}}, \citenamefont {{Giesler}},\ and\ \citenamefont
  {{Chen}}}]{2020PhRvD.102l3009Y}%
  \BibitemOpen
  \bibfield  {author} {\bibinfo {author} {\bibfnamefont {H.}~\bibnamefont
  {{Yu}}}, \bibinfo {author} {\bibfnamefont {S.}~\bibnamefont {{Ma}}}, \bibinfo
  {author} {\bibfnamefont {M.}~\bibnamefont {{Giesler}}},\ and\ \bibinfo
  {author} {\bibfnamefont {Y.}~\bibnamefont {{Chen}}},\ }\href
  {https://doi.org/10.1103/PhysRevD.102.123009} {\bibfield  {journal} {\bibinfo
   {journal} {\prd}\ }\textbf {\bibinfo {volume} {102}},\ \bibinfo {eid}
  {123009} (\bibinfo {year} {2020})},\ \Eprint
  {https://arxiv.org/abs/2007.12978} {arXiv:2007.12978 [gr-qc]} \BibitemShut
  {NoStop}%
\bibitem [{\citenamefont {{Kalogera}}(2000)}]{2000ApJ...541..319K}%
  \BibitemOpen
  \bibfield  {author} {\bibinfo {author} {\bibfnamefont {V.}~\bibnamefont
  {{Kalogera}}},\ }\href {https://doi.org/10.1086/309400} {\bibfield  {journal}
  {\bibinfo  {journal} {\apj}\ }\textbf {\bibinfo {volume} {541}},\ \bibinfo
  {pages} {319} (\bibinfo {year} {2000})},\ \Eprint
  {https://arxiv.org/abs/astro-ph/9911417} {arXiv:astro-ph/9911417 [astro-ph]}
  \BibitemShut {NoStop}%
\bibitem [{\citenamefont {{Belczynski}}\ \emph {et~al.}(2008)\citenamefont
  {{Belczynski}}, \citenamefont {{Taam}}, \citenamefont {{Rantsiou}},\ and\
  \citenamefont {{van der Sluys}}}]{2008ApJ...682..474B}%
  \BibitemOpen
  \bibfield  {author} {\bibinfo {author} {\bibfnamefont {K.}~\bibnamefont
  {{Belczynski}}}, \bibinfo {author} {\bibfnamefont {R.~E.}\ \bibnamefont
  {{Taam}}}, \bibinfo {author} {\bibfnamefont {E.}~\bibnamefont {{Rantsiou}}},\
  and\ \bibinfo {author} {\bibfnamefont {M.}~\bibnamefont {{van der Sluys}}},\
  }\href {https://doi.org/10.1086/589609} {\bibfield  {journal} {\bibinfo
  {journal} {\apj}\ }\textbf {\bibinfo {volume} {682}},\ \bibinfo {pages} {474}
  (\bibinfo {year} {2008})},\ \Eprint {https://arxiv.org/abs/astro-ph/0703131}
  {arXiv:astro-ph/0703131 [astro-ph]} \BibitemShut {NoStop}%
\bibitem [{\citenamefont {{Gerosa}}\ \emph {et~al.}(2013)\citenamefont
  {{Gerosa}}, \citenamefont {{Kesden}}, \citenamefont {{Berti}}, \citenamefont
  {{O'Shaughnessy}},\ and\ \citenamefont {{Sperhake}}}]{2013PhRvD..87j4028G}%
  \BibitemOpen
  \bibfield  {author} {\bibinfo {author} {\bibfnamefont {D.}~\bibnamefont
  {{Gerosa}}}, \bibinfo {author} {\bibfnamefont {M.}~\bibnamefont {{Kesden}}},
  \bibinfo {author} {\bibfnamefont {E.}~\bibnamefont {{Berti}}}, \bibinfo
  {author} {\bibfnamefont {R.}~\bibnamefont {{O'Shaughnessy}}},\ and\ \bibinfo
  {author} {\bibfnamefont {U.}~\bibnamefont {{Sperhake}}},\ }\href
  {https://doi.org/10.1103/PhysRevD.87.104028} {\bibfield  {journal} {\bibinfo
  {journal} {\prd}\ }\textbf {\bibinfo {volume} {87}},\ \bibinfo {eid} {104028}
  (\bibinfo {year} {2013})},\ \Eprint {https://arxiv.org/abs/1302.4442}
  {arXiv:1302.4442 [gr-qc]} \BibitemShut {NoStop}%
\bibitem [{\citenamefont {{Abbott}}\ \emph
  {et~al.}(2021{\natexlab{c}})\citenamefont {{Abbott}} \emph
  {et~al.}}]{2021arXiv211103606T}%
  \BibitemOpen
  \bibfield  {author} {\bibinfo {author} {\bibfnamefont {R.}~\bibnamefont
  {{Abbott}}} \emph {et~al.} (\bibinfo {collaboration} {LIGO, Virgo, and KAGRA
  Collaboration}),\ }\href@noop {} {\bibfield  {journal} {\bibinfo  {journal}
  {{}}\ } (\bibinfo {year} {2021}{\natexlab{c}})},\ \Eprint
  {https://arxiv.org/abs/2111.03606} {arXiv:2111.03606 [gr-qc]} \BibitemShut
  {NoStop}%
\bibitem [{\citenamefont {{Abbott}}\ \emph
  {et~al.}(2021{\natexlab{d}})\citenamefont {{Abbott}} \emph
  {et~al.}}]{2021arXiv211103634T}%
  \BibitemOpen
  \bibfield  {author} {\bibinfo {author} {\bibfnamefont {R.}~\bibnamefont
  {{Abbott}}} \emph {et~al.} (\bibinfo {collaboration} {LIGO, Virgo, and KAGRA
  Collaboration}),\ }\href@noop {} {\bibfield  {journal} {\bibinfo  {journal}
  {{}}\ } (\bibinfo {year} {2021}{\natexlab{d}})},\ \Eprint
  {https://arxiv.org/abs/2111.03634} {arXiv:2111.03634 [astro-ph.HE]}
  \BibitemShut {NoStop}%
\end{thebibliography}%

\end{document}